\documentclass[conference, final, compsoc, oneside, twocolumn, a4paper, 10pt]{IEEEtran}
\usepackage{cite}
\usepackage{amsmath,amssymb,amsfonts}
\usepackage{algorithmic}
\usepackage{graphicx}
\usepackage{textcomp}
\usepackage{xcolor}
\def\BibTeX{{\rm B\kern-.05em{\sc i\kern-.025em b}\kern-.08em
    T\kern-.1667em\lower.7ex\hbox{E}\kern-.125emX}}

\makeatletter

\def\ps@IEEEtitlepagestyle{%
	\def\@oddfoot{\mysubmissionnotice}%
	\def\@evenfoot{}%
}
\def\mysubmissionnotice{%
	{\footnotesize  Accepted (21.02.2020) for presentation at the 21st IEEE/ITG-Symposium on Photonic Networks, Leipzig, Germany, 13-14.05.2020.\hfill}
	\gdef\mysubmissionnotice{}
}

\makeatother

\usepackage[utf8]{inputenc}

\usepackage{tikz}

\usetikzlibrary{arrows}
\usetikzlibrary{arrows.meta}
\usetikzlibrary{spy}
\usetikzlibrary{chains}
\usetikzlibrary{positioning}
\usetikzlibrary{patterns}
\usetikzlibrary{calc}
\usetikzlibrary{backgrounds}
\usetikzlibrary{external}
\tikzstyle{block} = [draw, thick, rectangle, minimum height = 3em, minimum width = 3em, inner sep = 5pt, fill=white]
\tikzstyle{colorblock}= [draw, thick, rectangle, minimum height = 3em, minimum width = 3em, inner sep = 5pt, fill=olive]
\tikzstyle{source} = [draw, thick, circle, minimum size = 4em, inner sep = 0pt, fill=white]
\tikzstyle{sink} = [draw, thick, circle, minimum size = 4em, inner sep = 0pt, fill=white]
\tikzstyle{op} = [draw, circle, fill=white]
\tikzstyle{input} = [draw, circle, minimum size = 2mm, inner sep = 0pt, fill=white]
\tikzstyle{bullet} = [draw, circle, minimum size = 2mm, inner sep = 0pt, fill]
\tikzstyle{output} = [coordinate]
\tikzstyle{none} = 	[]

\usepackage{pgfplots}
\usepackage{pgfplotstable}
\usepackage{booktabs}
\usepgfplotslibrary{external} 
\usepgfplotslibrary{colormaps}  

\usepackage{siunitx}  
\usepackage{xfrac}

\usepackage[nolist]{acronym} 

\usepackage{caption}
\usepackage{subcaption}

\usepackage[above]{placeins}
\usepackage{float}
\usepackage{afterpage}

\begin{acronym}
 \acro{CSI}{channel state information}
 \acro{UE}{user equipment}
 \acro{UL}{uplink}
 \acro{BS}{basestation}
 \acro{TDD}{time division duplex}
 \acro{FDD}{frequency division duplex}
 \acro{ECC}{error-correcting code}
 \acro{MLD}{maximum likelihood decoding}
 \acro{HDD}{hard decision decoding}
 \acro{IF}{intermediate frequency}
 \acro{RF}{radio frequency}
 \acro{SDD}{soft decision decoding}
 \acro{NND}{neural network decoding}
 \acro{CNN}{convolutional neural network}
 \acro{ML}{maximum likelihood}
 \acro{GPU}{graphical processing unit}
 \acro{BP}{belief propagation}
 \acro{LTE}{Long Term Evolution}
 \acro{BER}{bit error rate}
 \acro{SNR}{signal-to-noise-ratio}
 \acro{ReLU}{rectified linear unit}
 \acro{BPSK}{binary phase shift keying}
 \acro{QPSK}{quadrature phase shift keying}
 \acro{AWGN}{additive white Gaussian noise}
 \acro{MSE}{mean squared error}
 \acro{LLR}{log-likelihood ratio}
 \acro{MAP}{maximum a posteriori}
 \acro{NVE}{normalized validation error}
 \acro{BCE}{binary cross-entropy}
 \acro{BLER}{block error rate}
 \acro{SQR}{signal-to-quantisation-noise-ratio}
 \acro{MIMO}{multiple-input multiple-output}
 \acro{OFDM}{orthogonal frequency division multiplex}
 \acro{RF}{radio frequency}
 \acro{LOS}{line of sight}
 \acro{NLoS}{non-line of sight}
 \acro{NMSE}{normalized mean squared error}
 \acro{CFO}{carrier frequency offset}
 \acro{SFO}{sampling frequency offset}
 \acro{IPS}{indoor positioning system}
 \acro{TRIPS}{time-reversal IPS}
 \acro{RSSI}{received signal strength indicator}
 \acro{MIMO}{multiple-input multiple-output}
 \acro{ENoB}{effective number of bits}
 \acro{AGC}{automated gain control}
 \acro{ADC}{analog to digital converter}
 \acro{ADCs}{analog to digital converters}
 \acro{FB}{front bandpass}
 \acro{FPGA}{field programmable gate array}
 \acro{JSDM}{Joint Spatial Division and Multiplexing}
 \acro{NN}{neural network}
 \acro{IF}{intermediate frequency}
 \acro{LoS}{line-of-sight}
 \acro{NLoS}{non-line-of-sight}
 \acro{DSP}{digital signal processing}
 \acro{AFE}{analog front end}
 \acro{SQNR}{signal-to-quantisation-noise-ratio}
 \acro{SINR}{signal-to-interference-noise-ratio}
 \acro{ENoB}{effective number of bits}
 \acro{AGC}{automated gain control}
 \acro{PCB}{printed circuit board}
 \acro{EVM}{error vector mangnitude}
 \acro{CDF}{cumulative distribution function}
 \acro{MRC}{maximum ratio combining}
 \acro{MRP}{maximum ratio precoding}
 \acro{MRT}{maximum ratio transmission}
 \acro{DeepL}{deep-learning}
 \acro{DL}{downlink}
 \acro{SISO}{single-input single-output}
 \acro{SGD}{stochastic gradient descent}
 \acro{CP}{cyclic prefix}
 \acro{MISO}{Multiple Input Single Output}
 \acro{LMMSE}{linear minimum mean square error}
 \acro{ZF}{zero forcing}
 \acro{USRP}{universal software radio peripheral}
 \acro{RNN}{recurrent neural network}
 \acro{GRU}{gated recurrent unit}
 \acro{LSTM}{long short-term memory}
 \acro{NTM}{neural turing machine}
 \acro{DNC}{differentiable neural computer}
 \acro{TCN}{temporal convolutional network}
 \acro{FCL}{fully connected layer}
 \acro{MANN}{memory augmented neural network}
 \acro{RNN}{recurrent neural network}
 \acro{SE}{spectral efficiency}
 \acro{CD}{chromatic dispersion}
 \acro{FIR}{finite impulse response}
 \acro{TBP}{time-bandwidth-product}
 \acro{AE}{autoencoder}
 \acro{SSFM}{split-step Fourier Method}
 \acro{KNL}{Kerr-nonlinearity}
 \acro{QAM}{quadrature-amplitude-modulation}
 \acro{DNN}{dense neural network}
 \acro{PSD}{power spectral density}
 \acro{ASE}{amplified spontaneous emission}
 \acro{NLSE}{nonlinear Schr\"odinger equation}
 \acro{SSFM}{split-step Fourier method}
 \acro{DAC}{digital to analog converter}
 \acro{LPF}{lowpass filter}
 \acro{TX}{transmitter}
 \acro{RX}{receiver}
 \acro{IQ}{in-phase-quadrature}
 \acro{DC}{direct-current}
 \acro{MI}{mutual information}
 \acro{PMF}{probability mass function}
\end{acronym}

\definecolor{mittelblau}{RGB}{0, 126, 198}
\definecolor{violettblau}{cmyk}{0.9, 0.6, 0, 0}
\definecolor{rot}{RGB}{238, 28 35}
\definecolor{apfelgruen}{RGB}{140, 198, 62}
\definecolor{gelb}{RGB}{255, 221, 0}
\definecolor{orange}{RGB}{244, 111, 33}
\definecolor{pink}{RGB}{237, 0, 140}
\definecolor{lila}{RGB}{128, 10, 145}
\definecolor{hellgrau}{RGB}{224, 224, 224}
\definecolor{mittelgrau}{RGB}{128, 128, 128}
\definecolor{dunkelgrau}{RGB}{80,80,80}
\definecolor{anthrazit}{RGB}{19, 31, 31}

\begin{document}

\title{Deep-learning Autoencoder for Coherent and Nonlinear Optical Communication\\
\thanks{This work has been supported by DFG, Germany, under grant BR 3205/6-1.}
}

\author{\IEEEauthorblockN{Tim Uhlemann, Sebastian Cammerer, Alexander Span, Sebastian D\"orner, and Stephan ten Brink\\}
	
\IEEEauthorblockA{\textit{Institute of Telecommunications} \\
\textit{University of  Stuttgart}\\
Pfaffenwaldring 47, 70569 Stuttgart, Germany\\
\{uhlemann,cammerer,span,doerner,tenbrink\}@inue.uni-stuttgart.de}

}

\maketitle

\begin{abstract}
Motivated by the recent success of end-to-end training of communications in the wireless domain, we strive to adapt the end-to-end-learning idea from the wireless case (i.e., linear) to coherent optical fiber links (i.e., nonlinear). Although, at first glance, it sounds like a straightforward extension, it turns out that several pitfalls exist – in terms of theory but also in terms of practical implementation. This paper analyzes an autoencoder’s potential and limitations for the optical fiber under the influence of Kerr-nonlinearity and chromatic dispersion. As there is no exact capacity limit known and, hence, no analytical perfect system solution available, we set great value to the interpretability on the learnings of the autoencoder. Therefore, we design its architecture to be as close as possible to the structure of a classic communication system, knowing that this may limit its degree of freedom and, thus, its performance. Nevertheless, we were able to achieve an unexpected high gain in terms of spectral efficiency compared to a conventional reference system.
\end{abstract}


\newcommand{\results}{../../optics/python/nb/tim/results/}
\newcommand{\classicawgn}{20200610105748}
\newcommand{\classiccd}{20200610102453}
\newcommand{\classiccdnl}{20200610092111}
\newcommand{\aeawgn}{20200610133403}
\newcommand{\aecd}{20200618074748}
\newcommand{\aecdnlzero}{20200619061001}
\newcommand{\aecdnlten}{20200624091812}
\newcommand{\aecdnltwenty}{20200620154530}
\newcommand{\aecdnlfourty}{20200621101723}

\newcommand{\ub}{\results\classiccdnl/csv/-10dBm/ub0.csv}
\newcommand{\lb}{\results\classiccdnl/csv/-10dBm/lb0.csv}
\newcommand{\classicawgnse}{\results\classicawgn/csv/10dBm/se.csv}
\newcommand{\classiccdse}{\results\classiccd/csv/10dBm/se.csv}
\newcommand{\classiccdnlse}{\results\classiccdnl/csv/-10dBm/se.csv}
\newcommand{\aeawgnse}{\results\aeawgn/csv/10dBm/se.csv}
\newcommand{\aecdse}{\results\aecd/csv/10dBm/se.csv}
\newcommand{\aecdnlzerose}{\results\aecdnlzero/csv/-10dBm/se.csv}
\newcommand{\aecdnltense}{\results\aecdnlten/csv/-10dBm/se.csv}
\newcommand{\aecdnltwentyse}{\results\aecdnltwenty/csv/-10dBm/se.csv}
\newcommand{\aecdnlfourtyse}{\results\aecdnlfourty/csv/-10dBm/se.csv}

\newcommand{\spectruminlpfhigh}{\results\classiccdnl/csv/-10dbm/spectrum_in,lpf.csv}
\newcommand{\spectrumoutrawlow}{\results\classiccdnl/csv/-10dbm/spectrum_out,raw.csv}
\newcommand{\spectrumoutrawmid}{\results\classiccdnl/csv/0dbm/spectrum_out,raw.csv}
\newcommand{\spectrumoutrawhigh}{\results\classiccdnl/csv/10dbm/spectrum_out,raw.csv}
\newcommand{\spectrumoutlpfhigh}{\results\classiccdnl/csv/10dbm/spectrum_out,lpf.csv}

\newcommand{\sinczerolow}{\results\aecdnlzero/csv/-10dBm/sinc.csv}
\newcommand{\sinczerolowphase}{\results\aecdnlzero/csv/-10dBm/sincphase.csv}
\newcommand{\shaperfilteredzerolow}{\results\aecdnlzero/csv/-10dBm/shaperfiltered.csv}
\newcommand{\decoderinputzerolow}{\results\aecdnlzero/csv/-10dBm/decoderinput.csv}
\newcommand{\shaperfilteredzerohigh}{\results\aecdnlzero/csv/0dBm/shaperfiltered.csv}
\newcommand{\decoderinputzerohigh}{\results\aecdnlzero/csv/0dBm/decoderinput.csv}
\newcommand{\shaperfilteredzeromax}{\results\aecdnlzero/csv/10dBm/shaperfiltered.csv}
\newcommand{\decoderinputzeromax}{\results\aecdnlzero/csv/10dBm/decoderinput.csv}
\newcommand{\shaperfilteredzerolowphase}{\results\aecdnlzero/csv/-10dBm/shaperfilteredphase.csv}
\newcommand{\decoderinputzerolowphase}{\results\aecdnlzero/csv/-10dBm/decoderinputphase.csv}
\newcommand{\shaperfilteredzerohighphase}{\results\aecdnlzero/csv/0dBm/shaperfilteredphase.csv}
\newcommand{\decoderinputzerohighphase}{\results\aecdnlzero/csv/0dBm/decoderinputphase.csv}
\newcommand{\shaperfilteredzeromaxphase}{\results\aecdnlzero/csv/10dBm/shaperfilteredphase.csv}
\newcommand{\decoderinputzeromaxphase}{\results\aecdnlzero/csv/10dBm/decoderinputphase.csv}
\newcommand{\embhigh}{\results\aecdnlzero/csv/-10dBm/emb.csv}
\newcommand{\shaperfilteredtwentylow}{\results\aecdnltwenty/csv/-10dBm/shaperfiltered.csv}
\newcommand{\decoderinputtwentylow}{\results\aecdnltwenty/csv/-10dBm/decoderinput.csv}
\newcommand{\shaperfilteredtwentyhigh}{\results\aecdnltwenty/csv/0dBm/shaperfiltered.csv}
\newcommand{\decoderinputtwentyhigh}{\results\aecdnltwenty/csv/0dBm/decoderinput.csv}
\newcommand{\shaperfilteredtwentymax}{\results\aecdnltwenty/csv/10dBm/shaperfiltered.csv}
\newcommand{\decoderinputtwentymax}{\results\aecdnltwenty/csv/10dBm/decoderinput.csv}
\newcommand{\shaperfilteredtwentylowphase}{\results\aecdnltwenty/csv/-10dBm/shaperfilteredphase.csv}
\newcommand{\decoderinputtwentylowphase}{\results\aecdnltwenty/csv/-10dBm/decoderinputphase.csv}
\newcommand{\shaperfilteredtwentyhighphase}{\results\aecdnltwenty/csv/0dBm/shaperfilteredphase.csv}
\newcommand{\decoderinputtwentyhighphase}{\results\aecdnltwenty/csv/0dBm/decoderinputphase.csv}
\newcommand{\shaperfilteredtwentymaxphase}{\results\aecdnltwenty/csv/10dBm/shaperfilteredphase.csv}
\newcommand{\decoderinputtwentymaxphase}{\results\aecdnltwenty/csv/10dBm/decoderinputphase.csv}
\newcommand{\spectruminlpfhighae}{\results\aecdnltwenty/csv/-10dbm/spectrum_in,lpf.csv}
\newcommand{\spectrumoutrawlowae}{\results\aecdnltwenty/csv/-10dbm/spectrum_out,raw.csv}
\newcommand{\spectrumoutrawmidae}{\results\aecdnltwenty/csv/0dbm/spectrum_out,raw.csv}
\newcommand{\spectrumoutrawhighae}{\results\aecdnltwenty/csv/10dbm/spectrum_out,raw.csv}
\newcommand{\spectrumoutlpfhighae}{\results\aecdnltwenty/csv/-10dbm/spectrum_out,lpf.csv}

\begin{figure*}
	\centering
	\scalebox{0.75}{
		\includegraphics{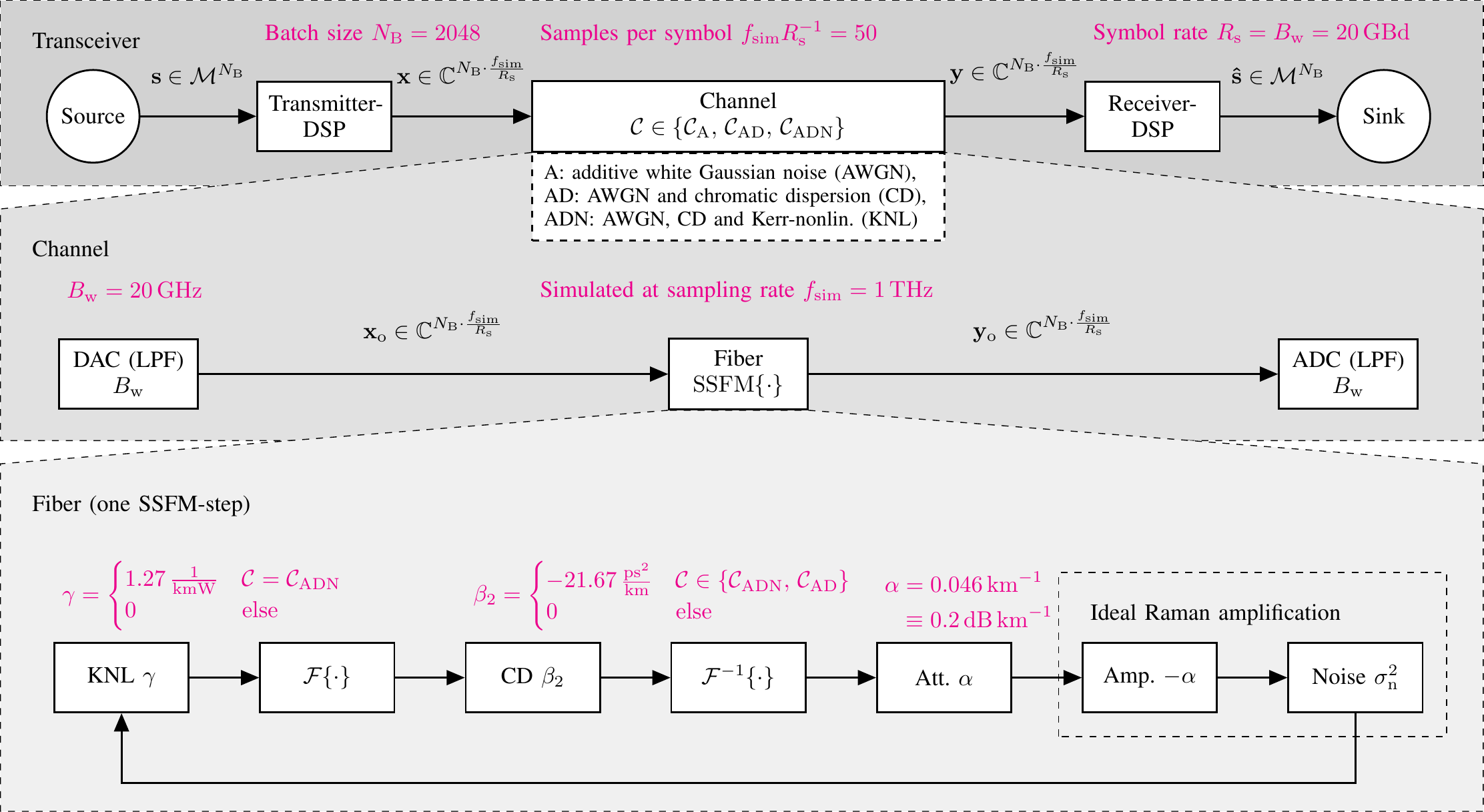}
	}
	\caption{Block-diagram describing the whole simulation setup including transmitter, channel and receiver. First, transmitter and receiver are conventionally implemented to obtain a baseline, and later replaced by a trainable AE.}
	\label{fig:system}
\end{figure*}

\section{Introduction}

At first glance, optical fibers promise the wistful dream of (wireless) communications engineers -- a transmission medium with seemingly infinite bandwidth, static propagation environments combined with low noise and small attenuation coefficients.
However, in the previous decades, the exponential increase of data-rates and the fast progress in circuitry have pushed the occupied bandwidth and sampling rates towards an operation point, where even the seemingly perfect optical fiber is dominated by non-linearity that cannot be neglected nor compensated easily anymore. As such, the optical fiber offers nonlinearity by its nature and, from an engineering perspective, opens up an exciting field of research. %

It can be easily justified that an increase of the fiber length or a larger alphabet size requires higher average launch power at the transmitter, which, in turn, increases the influence of the Kerr-effect and, thereby, introduces a possible spectral broadening. Hence, the \ac{TBP} and, finally, also the \ac{SE} are again lowered, which requires for classical compensation algorithms (i.e., post-compensation), that most of the signal's bandwidth needs to be captured at the receiver's \ac{ADC}. Otherwise there is a huge loss of information which effectively limits the maximum launch power and distance for data-transmission.
To account for these effects and to further increase the \ac{SE} of optical communication links, current literature can be split into two different research avenues, either the \emph{linearization} (compensation) of the nonlinear effects (cf. \cite{savory_digital_2010}; examples are digital back-propagation and \ac{CD}-compensation via \ac{FIR} filtering) or by designing a system that allows these effects and inherently operates in the \emph{nonlinear} regime (e.g., solitons \cite{yousefi_information_2014}). While the first approach benefits from the availability of many well-understood algorithms in the linear domain but comes at the price of a high complexity and certain limitations, the later approach becomes mathematically challenging and is, not yet, competitive in terms of the achievable \ac{SE}.

On the other hand, deep learning-driven communications \cite{oshea_introduction_2017} has become a promising and active research topic, in particular, in the wireless domain \cite{dorner_deep_2018,jiang2019turbo}. It has been shown that end-to-end learning of transmitter and receiver in a joint manner allows to find new signal constellations and even waveforms for almost arbitrary channels that are not restricted to linear scenarios \cite{farsad2017detection}. First applications of end-to-end learning in optics have been proposed by Karanov et al. \cite{karanov_end_2018} who have shown that an \ac{AE} may also be applied to a dispersive optical channel in a short-haul setup. In contrast to this Li et al. \cite{li_achievable_2018} and Jones et al. \cite{jones2019end} have shown that an \ac{AE} is also capable of learning and communicating over a simplified long-haul channel with \ac{KNL} only, i.e., disregarding \ac{CD}. 

We are attracted by the challenges of nonlinear optical fibers and the engineering simplicity of the end-to-end learning framework and, hence, seek to adapt the learning concepts from the wireless domain to the optical fiber. 

The main contribution of this work is to apply the \ac{AE} to a channel that includes both \ac{CD} and \ac{KNL}. Thereby, we train the coherent mapping as well as the pulse-shaping. We implement the channel via the \ac{SSFM} as a fully differentiable Tensorflow model which needs to be carefully implemented to ensure numerical stability, to guarantee the gradient-flow and to keep the required memory complexity during training managable.
Note that, although the \emph{universal approximation theorem} \cite{hornik1989multilayer} justifies that a neural network can potentially approximate the optimal transceiver function, it does not state that the training process will practically converge towards a suitable solution. In other words, the training complexity becomes the practical limitation of learning-based systems \cite{gruber2017deep}. Thus, it requires to limit the degrees of freedom during training by providing a carefully adjusted AE structure. %

One of the key ideas is to impose an autoencoder-structure that preserves flexibility of a machine learning algorithm but allows the interpretation and comparison by means of classical communications. For this, we sequentially activate impairments such as \ac{AWGN}, \ac{CD} and \ac{KNL} while always comparing the achieved results to a \emph{conventional} baseline system as benchmark to understand its individual implications to the autoencoder-framework and to understand how well the autoencoder can compensate for these effects. Nevertheless, we have been able to achieve a high gain in terms of \ac{SE} at high average power.

In Section \ref{sec:system} we introduce the channel model. Section \ref{sec:classical} details the reference system and its challenges. The actual \ac{AE} is introduced in Section \ref{sec:ae}, followed by its results in Section \ref{sec:results}. Finally, we draw a conclusion in Section \ref{sec:conclusion}.

\begin{figure*}[t]
	\begin{subfigure}[c]{\textwidth}
		\centering
		\scalebox{0.8}{
			\includegraphics{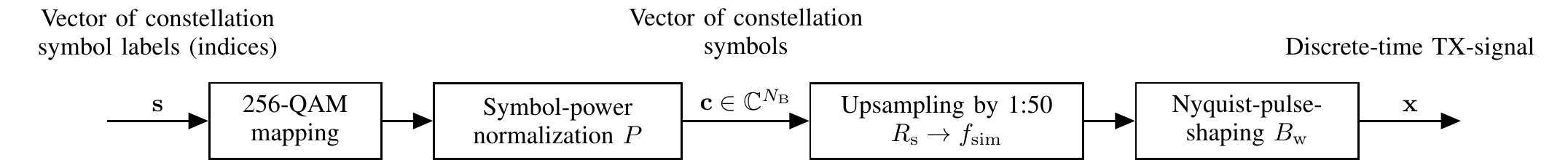}
		}
		\subcaption{Conventional \ac{TX}-\ac{DSP}.}
		\label{fig:classic-tx-dsp}
	\end{subfigure}
	\begin{subfigure}[c]{\textwidth}
		\centering
		\scalebox{0.8}{
			\includegraphics{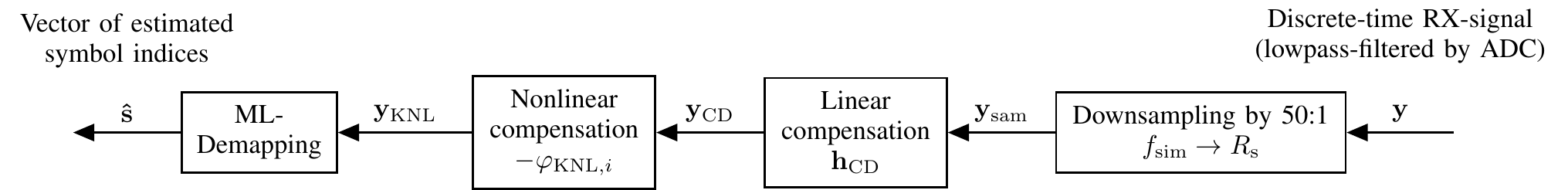}
		}
		\subcaption{Conventional \ac{RX}-\ac{DSP}.}
		\label{fig:classic-rx-dsp}
	\end{subfigure}
\caption{Blockdiagrams of the conventional (classic) DSP implementations.}
\label{fig:classic-dsp}
\end{figure*}

\section{Channel model}\label{sec:system}

The block diagram of the simulation setup is shown in Fig.~\ref{fig:system}. We consider a single polarization nonlinear optical fiber channel $\mathcal{C}=\mathcal{C}_\mathrm{ADN}$ with ideal distributed Raman amplification where the index denotes the included impairments \ac{AWGN} ($\mathrm{A}$), \ac{CD} ($\mathrm{D}$) and \ac{KNL} ($\mathrm{N}$), respectively. The propagation of an optical signal $q(t,z)$ along such a channel can be described by the \ac{NLSE}.
\begin{equation}
\frac{\partial q(t,z)}{\partial z}= j\frac{\beta_2}{2}\frac{\partial^2 q(t,z)}{\partial t^2}-j\gamma |q(t,z)|^2 q(t,z) + n(t,z)
\end{equation}
with $\beta_2$ and $\gamma$ being the dispersion and nonlinearity coefficient, respectively. Amplified spontaneous emission (\acs{ASE}) noise is represented by $n(t,z)$. This channel model accounts for both \ac{CD} and the \ac{KNL}, which both cause either temporal or spectral broadening. The fiber attenuation is assumed to be perfectly compensated by distributed amplification. For simulation, this channel model has been implemented using the symmetric \ac{SSFM}. It divides the fiber length $\ell$ into $N_\mathrm{SSFM}=\ell/\Delta z$ equal segments and solves the NLSE iteratively. In each iteration, \ac{KNL}, \ac{CD} and noise are treated separately.

The solution for the KNL within a step $\Delta z$ is given in time domain as
\begin{equation}
q(t,z+\Delta z)=q(t,z) \exp(-j \gamma |q(t,z)|^2 \Delta z).
\end{equation}
The effect of CD is accounted for in the frequency domain as
\begin{equation}
Q(\omega,z + \Delta z)=Q(\omega,z) \exp \left(-\frac{j}{2} \beta_2 \omega^2  \Delta z \right)
\end{equation}
where $Q(\omega,z)=\mathcal{F}\{q(t,z)\}$ is the Fourier transform of the optical signal. The ideal distributed Raman amplification induces noise with power spectral and spatial density 
\begin{equation}
\rho_\mathrm{n} = n_\mathrm{sp} h f_0 \alpha
\end{equation}
where $n_\mathrm{sp}$ is the spontaneous emission factor, $h$ is Planck's constant, $f_0$ is the carrier frequency and $\alpha$ denotes the fiber attenuation coefficient. For simulation, \ac{AWGN} with variance $\sigma_\mathrm{n}^2=\rho_\mathrm{n} \Delta z f_\mathrm{sim}$ is added to the signal in each step $\Delta z$, where $f_\mathrm{sim}$ is the simulation bandwidth (i.e., sampling rate) of the channel. 

We assume an ideal coherent transmitter and receiver where the bandwidth limitation $B_\mathrm{w}$ of both the \ac{DAC} and \ac{ADC} is taken into account by a respective \ac{LPF}. Note that the sampling rate is not changed by \ac{DAC} and \ac{ADC}, though being band-limited. This makes it easier to later compare the signals that are generated by \ac{TX} and signals at the channel output.

For modulation, the symbol alphabet size is chosen to be $M=256$ (hence, the number of bits per symbol is $M_\mathrm{b}=8\, \mathrm{\sfrac{bits}{symbol}}$) with a blockbased signal processing and transmission of $N_\mathrm{B}=2048$ symbols (batch size). Each batch is represented at the input of the transmitter by a vector of constellation symbol labels (indices) $\mathbf{s}\in\left\{0,1,\,2,\,\ldots,\,M-1\right\}^{N_\mathrm{B}}=\mathcal{M}^{N_\mathrm{B}}$. The transmission signal sequence is denoted as $\mathbf{x}\in\mathbb{C}^{N_\mathrm{B}\cdot \sfrac{f_\mathrm{sim}}{R_\mathrm{s}}}$ and has a symbol rate of $R_\mathrm{s}=B_\mathrm{w}=20\,\mathrm{GHz}$ (according to the \ac{DAC}-bandwidth and simulation sampling rate), i.e., the number of samples per symbol is $\sfrac{f_\mathrm{sim}}{R_\mathrm{s}} = \sfrac{f_\mathrm{sim}}{B_\mathrm{w}} \in \mathbb{N}$. The optical signal launched into the fiber is represented in simulation by $\mathbf{x}_\mathrm{o}\in\mathbb{C}^{N_\mathrm{B}\cdot \sfrac{f_\mathrm{sim}}{R_\mathrm{s}}}$. The signal sequence after propagation along the fiber is denoted as $\mathbf{y}_\mathrm{o}\in\mathbb{C}^{N_\mathrm{B}\cdot \sfrac{f_\mathrm{sim}}{R_\mathrm{s}}}$. The AD conversion at the receiver results in $\mathbf{y}\in\mathbb{C}^{N_\mathrm{B}\cdot \sfrac{f_\mathrm{sim}}{R_\mathrm{s}}}$. The recovered symbol index vector at the receiver is $\mathbf{\hat{s}} \in \mathcal{M}^{N_\mathrm{B}}$. The transmitter \ac{DSP} maps the symbol index vector $\mathbf{s}$ to the transmission signal $\mathbf{x}$. The receiver \ac{DSP} recovers a symbol index vector $\mathbf{\hat{s}}$ from the received $\mathbf{y}$. The transmitter and receiver \acp{DSP} are implemented by an \ac{AE} as described in Section~\ref{sec:ae}.

The channel and simulation parameters are given in Tab.~\ref{tab:channel}. Note that the simulation bandwidth needs to be chosen carefully in order to ensure proper simulation of the analog waveform channel. The signal bandwidth along the link is generally unknown due the potential bandwidth expansion. We approximate the maximum occurring bandwidth according to \cite{agrawal_nonlinear_2000} as $\hat{B}_\mathrm{max}=0.86 \cdot \gamma \left(2\cdot P_\mathrm{max}\right) \ell B_\mathrm{w}$, where $P_\mathrm{max}$ is the maximum average input power, which is used in the following sections. We choose $f_\mathrm{sim}=1\,\mathrm{THz}$; this way we have $f_\mathrm{sim}>2\hat{B}_\mathrm{max}$ in all scenarios.

We define the \ac{SNR} as $SNR=\sfrac{P}{\rho_\mathrm{n} \ell B_\mathrm{w}}$, i.e., the ratio of the mean input signal power and the noise power within the maximum bandwidth of the transmit signal. However, for a nonlinear fiber channel, the \ac{SNR} is not a sufficient quantity to characterize the system performance. The above \ac{SNR} definition does not account for nonlinear interaction of the signal with noise components outside the signal bandwidth.\footnote{This outside-bandwidth interaction may be interpreted as an additional noise term, that degrades the system performance.} Furthermore, the system performance does also explicitly depend on the absolute signal power. Nonetheless, in the linear regime of the fiber (low signal power), the \ac{SNR} allows to compare the system performance with references such as, e.g., the \ac{AWGN} channel. 

\begin{table}[h]
	\caption{Channel parameters}
	\begin{center}
		\begin{tabular}{c|c|c}
			\textbf{Property}&\textbf{Symbol}&\textbf{Value} \\
			\hline
			&& \\
			Planck's constant&$h$&$6.626\cdot 10^{-34}\,\si{\joule\second}$ \\
			Carrier frequency&$f_0$&$193.55\,\si{\tera\hertz}$ \\
			Attenuation&$\alpha$&$0.046\,\si{\per\kilo\meter} \equiv 0.2\, \mathrm{dB}\,\mathrm{km}^{-1}$ \\
			Chromatic dispersion&$\beta_2$&$-21.67\,\si{\pico\second\squared\per\kilo\meter}$ \\
			Kerr-nonlinearity&$\gamma$&$1.27\,\si{\per\kilo\meter\per\watt}$ \\
			Spontaneous emission&$n_\mathrm{sp}$&$1$ \\
			Fiber length&$\ell$&$1000\,\si{\kilo\meter}$ \\
			Simulation sampling rate&$f_\mathrm{sim}$&$1\,\si{\tera\hertz}$ \\
			No. of SSFM-steps&$N_\mathrm{SSFM}$&$ 200 $ \\
			Launch power&$P$&$\left[-30\,\mathrm{dBm},\,10\,\mathrm{dBm}\right]$ \\
			Bandwidth of TX/RX&$B_\mathrm{w}$&$20\,\si{\giga\hertz}$ \\
		\end{tabular}
		\label{tab:channel}
	\end{center}
\end{table}
\section{Conventional system and performance}\label{sec:classical}

Having discussed the channel, we can now introduce the remaining \ac{DSP}-blocks for transmitter and receiver. To validate the \ac{AE}-performance and to obtain a baseline we first implemented a basic and conventional \ac{DSP}-algorithm to simulate a data transmission and compensate for the impairments. We are aware of the fact that more sophisticated (but also more complex) algorithms exist. Nevertheless, this simple baseline helps to interpret the effect of each impairment and the respective compensation that is learned by the \ac{AE} in Section \ref{sec:ae}.

The conventional system's transmitter-\ac{DSP} (shown in Fig. \ref{fig:classic-tx-dsp}) consists of a 256-\ac{QAM}-Mapper, followed by a power normalization to obtain \ac{IQ}-symbols with the desired launch power $P$. After upsampling, a Nyquist-pulse-shaping is applied, to not exceed the required bandwidth $B_\mathrm{w}$ and avoid inter-symbol interference.

The corresponding receiver-\ac{DSP}, shown in Fig. \ref{fig:classic-rx-dsp}, performs a compensation of \ac{CD} and \ac{KNL} as follows \cite{savory_digital_2010}.
Concerning \ac{CD} we use an \ac{FIR}-filter $\mathbf{h}_\mathrm{CD}$ with taps
\begin{equation}
	\label{eq:cd-compensation}
	h_{\mathrm{CD},k} = \frac{1}{\sqrt{\nu}}\cdot \exp\left(-j\frac{\pi}{\nu}\left[k-\frac{N_\mathrm{CD}-1}{2}\right]^2\right)
\end{equation}
where $\nu=2\pi\beta_2 \ell B_\mathrm{w}^2$ and $N_\mathrm{CD}=\left\lfloor \left|\nu\right|\right\rfloor$; 
Eq. (\ref{eq:cd-compensation}) is convolved with the received signal, such that $\mathbf{y}_\mathrm{CD}~=~\mathbf{y}_\mathrm{sam}~*~\mathbf{h}_{\mathrm{CD}}$. Here, $\mathbf{y}_\mathrm{sam}$ is the downsampled received signal $\mathbf{y}$ with symbol rate $R_\mathrm{s}$. 
For the \ac{KNL}-compensation we have implemented a simple power-dependent and sample-wise back-rotation described as
\begin{equation}
	y_{\mathrm{KNL}, i}=y_{\mathrm{CD}, i}\cdot \exp\left(j\cdot \gamma \left| y_{\mathrm{CD}, i}\right|^2 \ell \right)
\end{equation}
where $i$ is the discrete-time-index of the signal and $\varphi_{\mathrm{KNL},i} = \gamma \left| y_{\mathrm{CD}, i}\right|^2 \ell$ is the nonlinear sample-wise phase-shift. Finally, a maximum likelihood demapping of the received and sampled signal leads to the estimated symbol index.

As a performance measure (to later compare the conventional system with the AE) we use the average \ac{MI} defined as
\begin{equation}
MI=H(\mathbf{s})+H(\mathbf{\hat{s}})-H(\mathbf{s},\,\mathbf{\hat{s}})
\end{equation}
where $H(\cdot)$ is the entropy and $H(\cdot,\,\cdot)$ the joint entropy. Note that $H(\mathbf{\hat{s}})=\log_{2}(p_{\hat{s}}(\xi))$, where the \ac{PMF} $p_{\hat{s}}(\xi)$ is obtained by generating the histogram over the hard decisions $\mathbf{\hat{s}}$ of the respective receiver-\ac{DSP}, and normalizing it appropriately. The same holds for the required joint distribution. Further, we use the derived \ac{SE} as
\begin{equation}
SE=\frac{MI}{T \cdot B_\mathrm{w}}
\end{equation}
where $T=\sfrac{1}{B_\mathrm{w}}$ is the symbol duration and $TBP = T\cdot B_\mathrm{w} = 1$ is denoted as the \acf{TBP}. It follows that $TBP=1\Rightarrow SE=MI$. The \ac{SE} over \ac{SNR} defined in Section \ref{sec:system} (or input power $P$, respectively) for the conventional system and the \ac{AE} shall be compared, together with corresponding reference curves and bounds. Therefore, Shannon's capacity-limit for the \ac{AWGN}-channel $\mathcal{C}_\mathrm{A}$, which also holds for the optical fiber \cite{kramer_upper_2015} and the symbol-wise capacity of a conventional 256-\ac{QAM} for $\mathcal{C}_\mathrm{A}$ shall be considered.

\begin{figure}
	\includegraphics{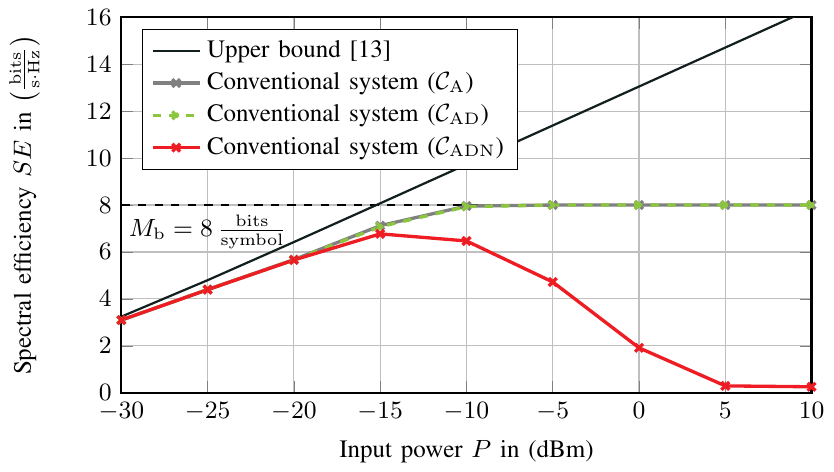}
	\caption{\ac{SE} over input power $P$ of the chosen \emph{conventional} system for $\mathcal{C}_\mathrm{AD}$ and $\mathcal{C}_\mathrm{ADN}$, depicting that \ac{CD} can be compensated completely, while additional \ac{KNL} can not.}
	\label{fig:classic-performance}
\end{figure}

\begin{figure}
	\includegraphics{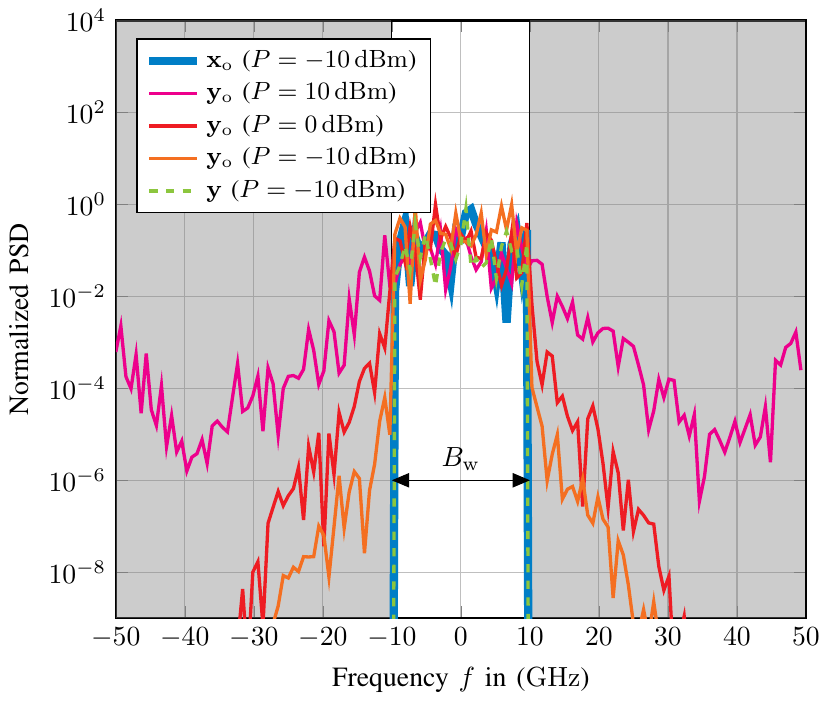}
	\caption{Estimated and peak normalized \ac{PSD} (using Welch's method) of the input and output signal before and after the \ac{ADC} for the \emph{conventional} system and different input powers. The bandwidth-expansion due to \ac{KNL} can clearly be seen.}
	\label{fig:psd-classic}
\end{figure}

Fig. \ref{fig:classic-performance} shows the reference performance for a simplified channel $\mathcal{C}_\mathrm{AD}$ (incl. \ac{CD}). It achieves the same performance as for the above spoken $\mathcal{C}_\mathrm{A}$ and hence, the conventional system completely compensates \ac{CD} by $\mathbf{h}_\mathrm{CD}$. The same holds for $\mathcal{C}_\mathrm{ADN}$ and low input powers $P$, where \ac{KNL} is not yet significant. As expected, the achieved performance increases with launch-power until it reaches its maximum and results in a drop due to \ac{KNL}.
Fig. \ref{fig:psd-classic} shows the \ac{PSD} of $\mathbf{x}$ after \ac{TX}-\ac{DSP}, $\mathbf{y}_\mathrm{o}$ before and  $\mathbf{y}$ after the channel's (or \ac{ADC}'s, respectively) \ac{LPF}. Here, one can see that the drop is most probably due to the \ac{KNL}-caused bandwidth-expansion of the signal at high input powers (e.g., for $P>0\,\mathrm{dBm}$). Note that, information falling outside of the \ac{DSP} baseband bandwidth $B_\mathrm{w}$ gets lost by the receiver's (i.e., the ADC's) lowpass characteristics! This motivates the setup of the chosen \ac{AE} that is supposed to find a pulse shape that propagates with less distortion through the fiber while keeping the information within the given bandwidth $B_\mathrm{w}$.

\begin{figure*}[t]
	\begin{subfigure}[c]{\textwidth}
		\centering
		\scalebox{0.8}{
			\includegraphics{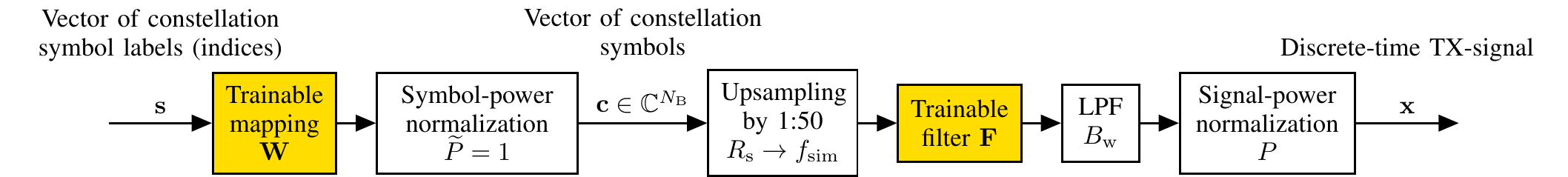}
		}
		\subcaption{\ac{AE}-\ac{TX}-\ac{DSP}.}
		\label{fig:ae-tx}
	\end{subfigure}
	\begin{subfigure}[c]{\textwidth}
		\centering
		\scalebox{0.8}{
			\includegraphics{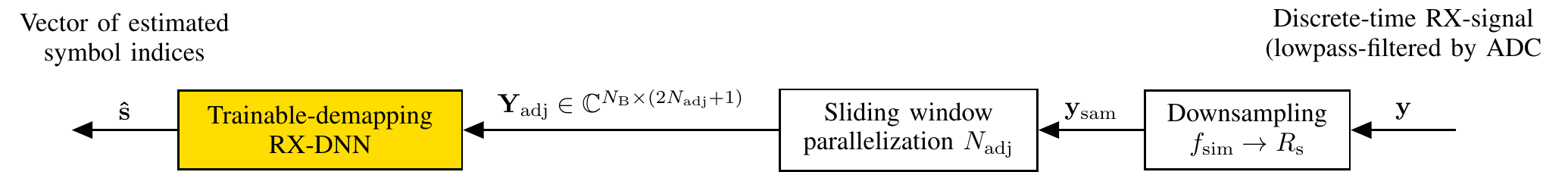}
		}
		\subcaption{\ac{AE}-\ac{RX}-\ac{DSP}.}
		\label{fig:ae-rx}
	\end{subfigure}
	\caption{Block-diagrams of the AE's DSP implementations. Trainable blocks are in yellow.}
	\label{fig:ae-dsp}
\end{figure*}

\begin{figure}
	\includegraphics{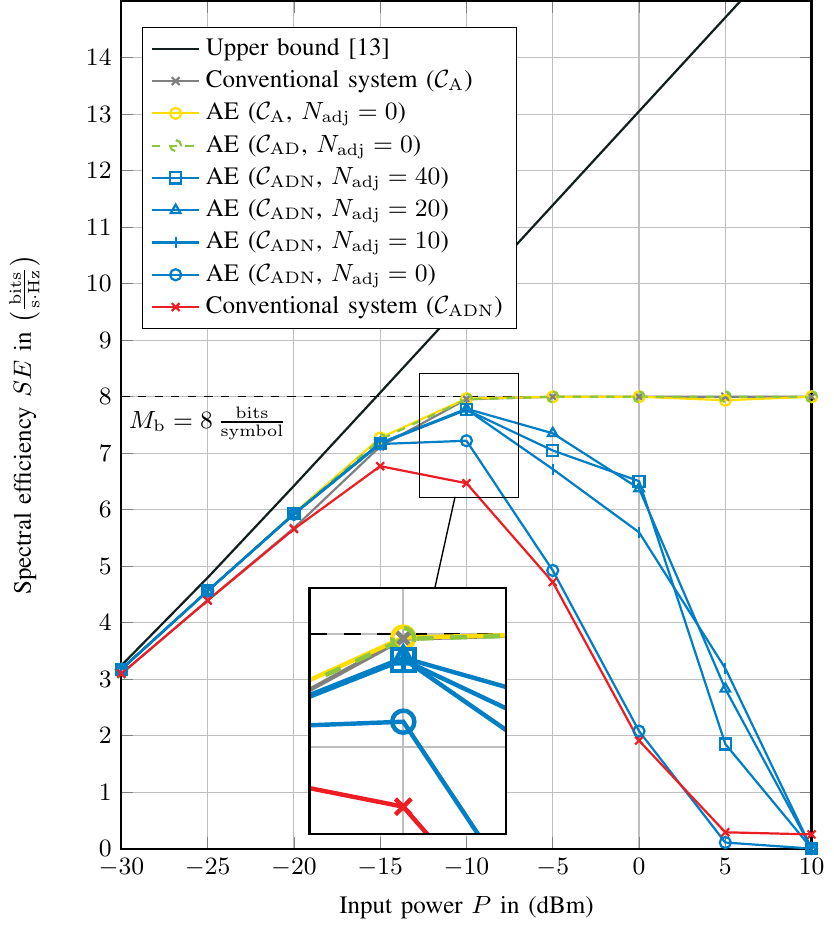}%
	\caption{\ac{SE} over input power $P$ of the \emph{\ac{AE}} for different channel models and compared with the chosen \emph{conventional} system.}
	\label{fig:ae-performance}
\end{figure}

\begin{figure}
	\includegraphics{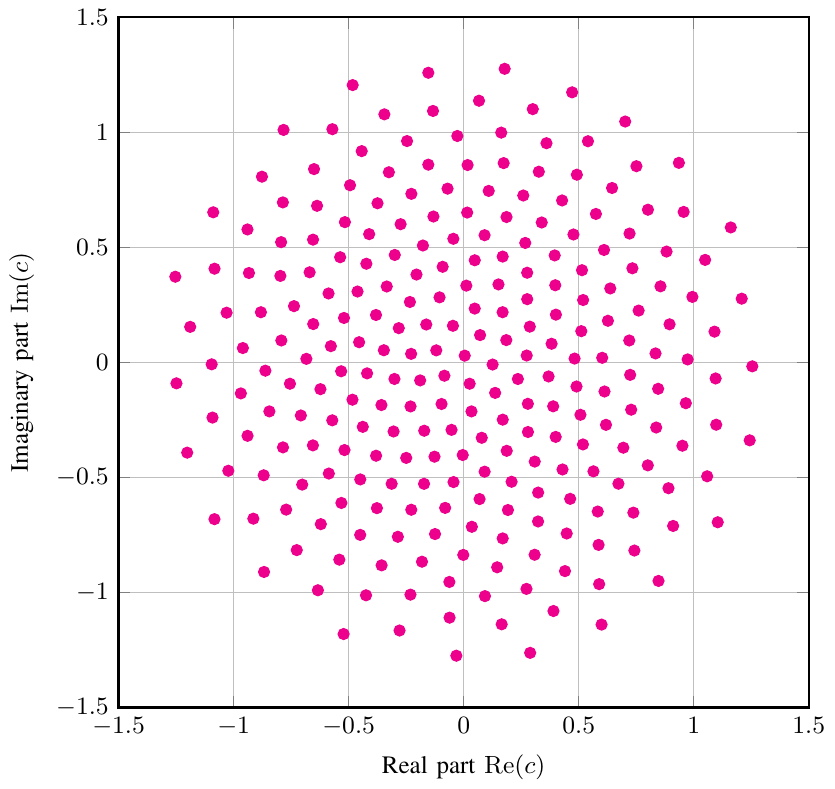}
	\caption{Constellation symbols $c$ learned by the AE, or, more precisely, the rows of its embedding matrix $\mathbf{W}$.}
	\label{fig:constellation}
\end{figure}

\begin{figure*}
	\begin{subfigure}[c]{0.5\textwidth}
		\includegraphics{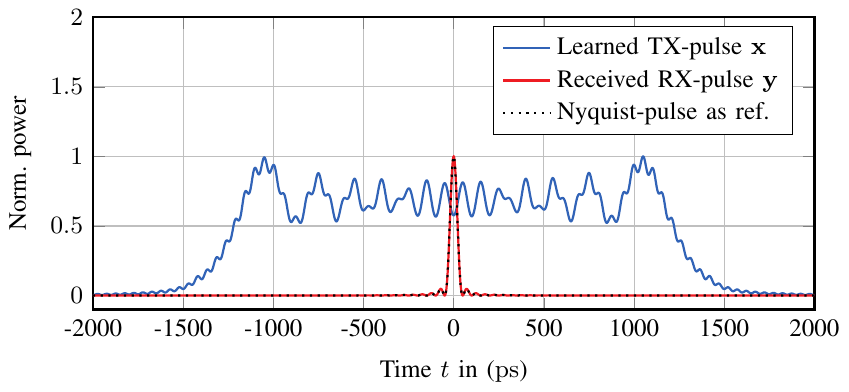}
		\vspace{-15pt}
		\subcaption{Normalized signal power at $P=-10\,\mathrm{dBm}$.}
		\vspace{5pt}
		\label{subfig:zero-amplitude-low}
	\end{subfigure}
	\begin{subfigure}[c]{0.5\textwidth}
		\includegraphics{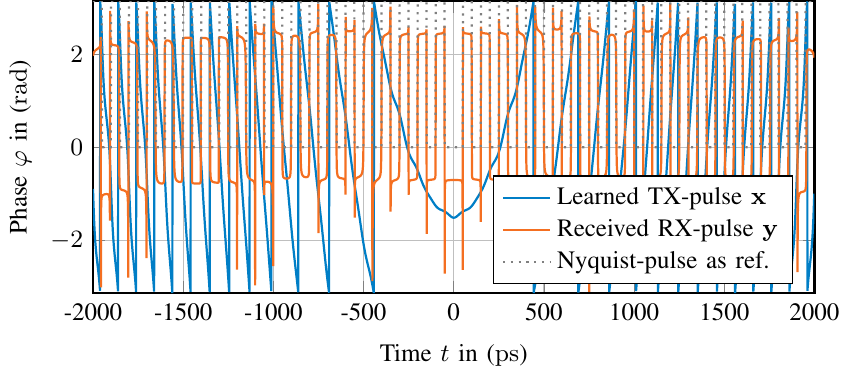}
		\vspace{-15pt}
		\subcaption{Phase at $P=-10\,\mathrm{dBm}$.}
		\vspace{5pt}
		\label{subfig:zero-phase-low}
	\end{subfigure}
	\begin{subfigure}[c]{0.5\textwidth}
		\includegraphics{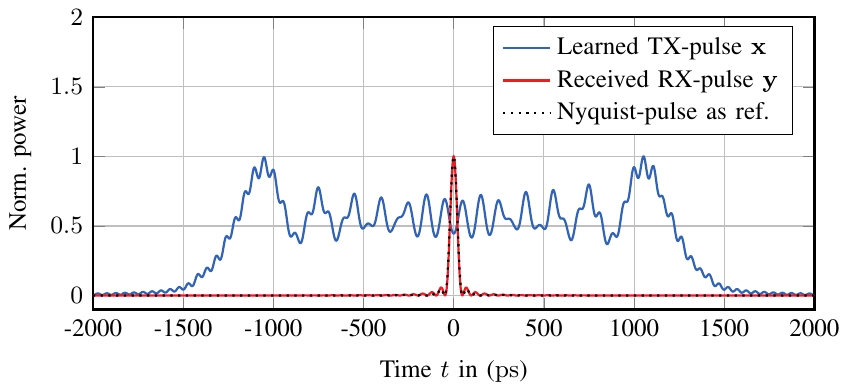}
		\vspace{-15pt}
		\subcaption{Normalized signal power at $P=0\,\mathrm{dBm}$.}
		\vspace{5pt}
		\label{subfig:zero-amplitude-mid}
	\end{subfigure}
	\begin{subfigure}[c]{0.5\textwidth}
		\includegraphics{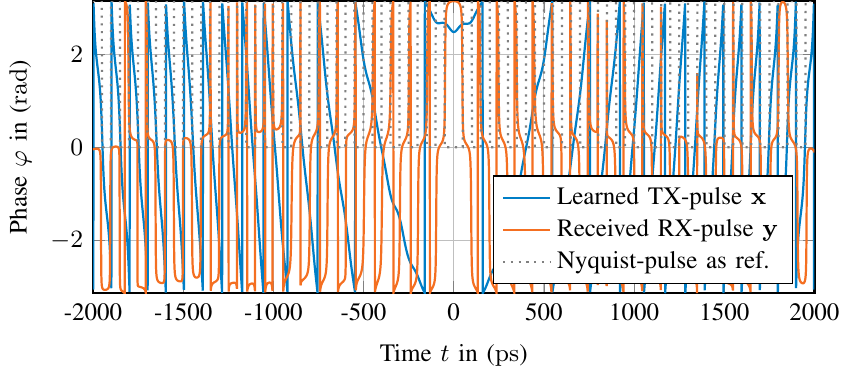}
		\vspace{-15pt}
		\subcaption{Phase at $P=0\,\mathrm{dBm}$.}
		\vspace{5pt}
		\label{subfig:zero-phase-mid}
	\end{subfigure}
	\begin{subfigure}[c]{0.5\textwidth}
		\includegraphics{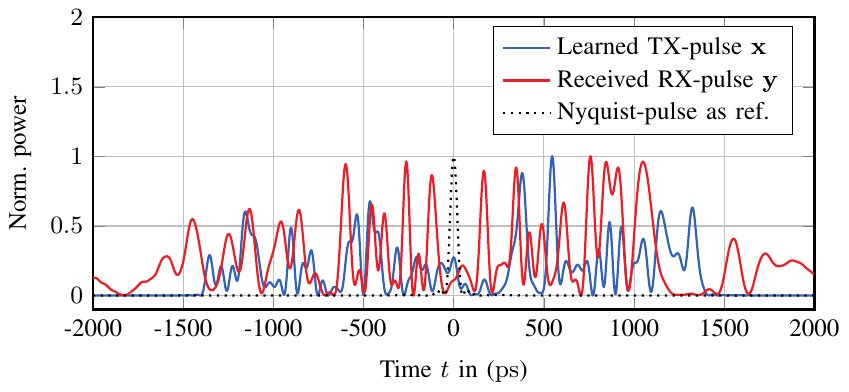}
		\vspace{-15pt}
		\subcaption{Normalized signal power at $P=10\,\mathrm{dBm}$.}
		\vspace{5pt}
		\label{subfig:zero-amplitude-high}
	\end{subfigure}
	\begin{subfigure}[c]{0.5\textwidth}
		\includegraphics{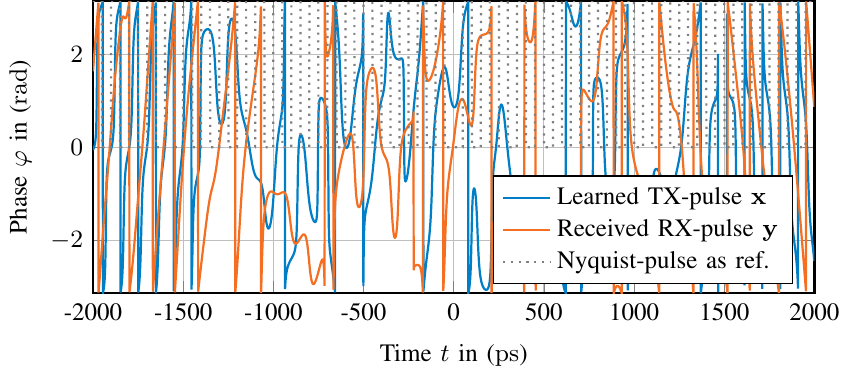}
		\vspace{-15pt}
		\subcaption{Phase at $P=10\,\mathrm{dBm}$.}
		\vspace{5pt}
		\label{subfig:zero-phase-high}
	\end{subfigure}
	\caption{Learned pulse shapes $\mathbf{x}$ (with $N_\mathrm{B}=1$) of the AE TX-DSP's trainable filter $\mathbf{F}$ with $N_\mathrm{adj}=0$ and the corresponding received pulse $\mathbf{y}$ at the RX-DSP. The Nyquist-pulse $\frac{\sin(\pi R_\mathrm{s} t)}{\pi R_\mathrm{s} t}$ is given as reference.}
	\label{fig:ae-pulse-shapes-zero}
\end{figure*}

\begin{figure}
	\includegraphics{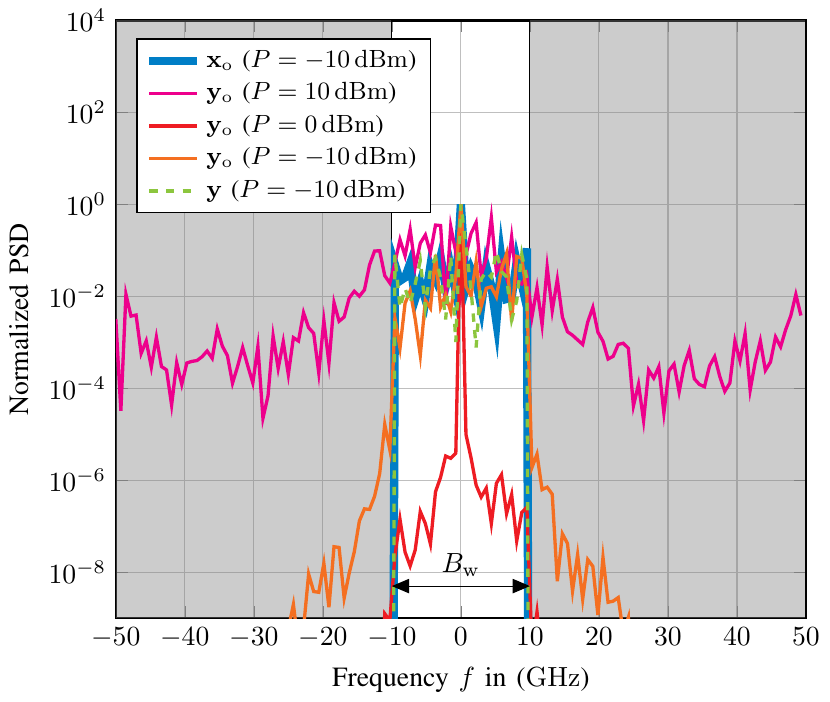}
	\caption{Estimated and peak normalized \acp{PSD} for the \emph{\ac{AE}}. Compared to the conventional system the AE achieved less spectral broadening for $P\leq 0\,\mathrm{dBm}$.}
	\label{fig:psd-ae}
\end{figure}

\begin{figure*}
	\begin{subfigure}[c]{0.5\textwidth}
		\includegraphics{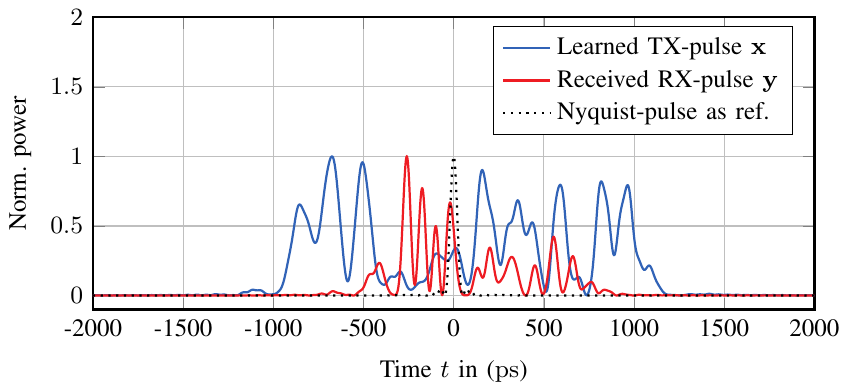}
		\vspace{-15pt}
		\subcaption{Normalized signal power at $P=-10\,\mathrm{dBm}$.}
		\vspace{5pt}
		\label{subfig:twenty-amplitude-low}
	\end{subfigure}
	\begin{subfigure}[c]{0.5\textwidth}
		\includegraphics{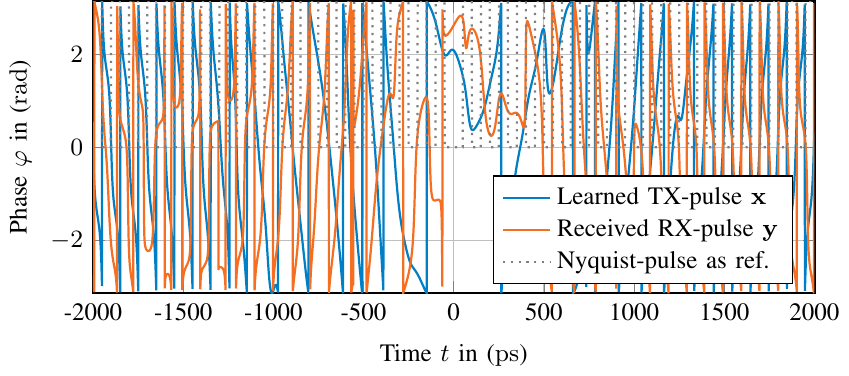}
		\vspace{-15pt}
		\subcaption{Phase at $P=-10\,\mathrm{dBm}$.}
		\vspace{5pt}
		\label{subfig:twenty-phase-low}
	\end{subfigure}
	\begin{subfigure}[c]{0.5\textwidth}
		\includegraphics{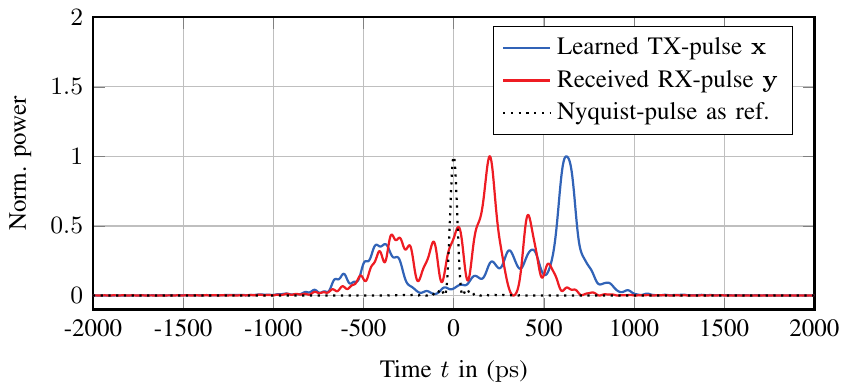}
		\vspace{-15pt}
		\subcaption{Normalized signal power at $P=0\,\mathrm{dBm}$.}
		\vspace{5pt}
		\label{subfig:twenty-amplitude-mid}
	\end{subfigure}
	\begin{subfigure}[c]{0.5\textwidth}
		\includegraphics{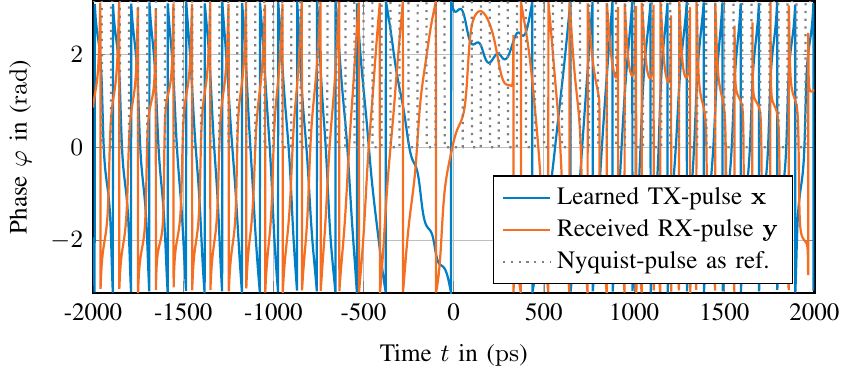}
		\vspace{-15pt}
		\subcaption{Phase at $P=0\,\mathrm{dBm}$.}
		\vspace{5pt}
		\label{subfig:twenty-phase-mid}
	\end{subfigure}
	\begin{subfigure}[c]{0.5\textwidth}
		\includegraphics{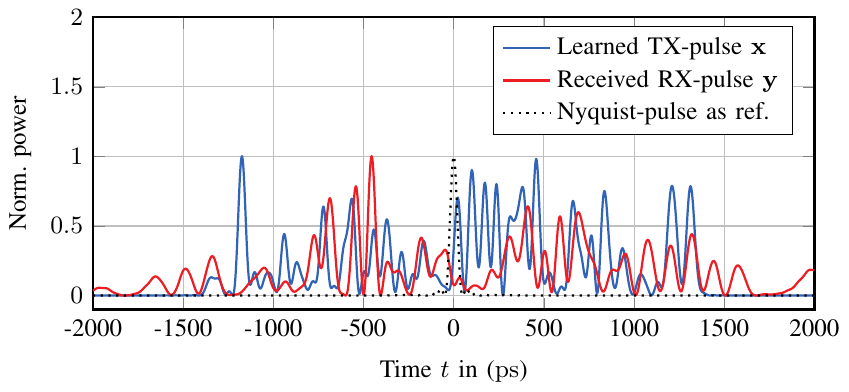}
		\vspace{-15pt}
		\subcaption{Normalized signal power at $P=10\,\mathrm{dBm}$.}
		\vspace{5pt}
		\label{subfig:twenty-amplitude-high}
	\end{subfigure}
	\begin{subfigure}[c]{0.5\textwidth}
		\includegraphics{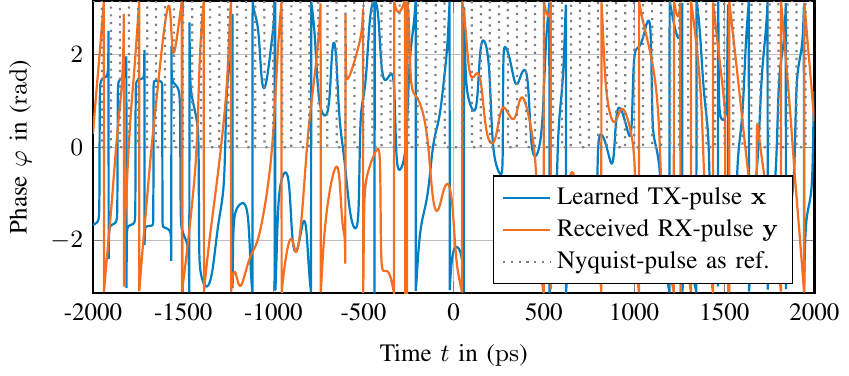}
		\vspace{-15pt}
		\subcaption{Phase at $P=10\,\mathrm{dBm}$.}
		\label{subfig:twenty-phase-high}
	\end{subfigure}
	\caption{Learned pulse shapes $\mathbf{x}$ (with $N_\mathrm{B}=1$) of the AE TX-DSP's trainable filter $\mathbf{F}$ with $N_\mathrm{adj}=20$ and the corresponding received pulse $\mathbf{y}$ at the RX-DSP. The Nyquist-pulse $\frac{\sin(\pi R_\mathrm{s} t)}{\pi R_\mathrm{s} t}$ is given as reference.}
	\label{fig:ae-pulse-shapes-twenty}
\end{figure*}

\section{Autoencoder}\label{sec:ae}

Usually, an \ac{AE} in the context of communications consists of two neural networks (encoder and decoder) with a penalty layer in between, which, in our case, is the optical channel. Its goal is to encode the input information, such that it can be decoded with minimal loss after having passed the penalty. One problem concerning this approach is that the encoding network is hard to interpret. Hence, we have designed an architecture for the \ac{AE} that follows along the lines of classic communication, meaning that it has a classic communication structure while specific parameters may be learned. Those trainable parameters are
\begin{itemize}
	\item the mapping of symbol indices $\mathbf{s}$ to \ac{IQ}-symbols $\mathbf{c}$, i.e., embedding $\mathbf{W}$, and
	\item the pulse-shaping FIR-filter, i.e., its filter-taps $\mathbf{F}$
\end{itemize}
Fig. \ref{fig:ae-tx} shows the trainable \ac{TX}-\ac{DSP}-implementation of the \ac{AE}. Akin to a conventional system, first, it maps the integer symbol indices $\mathbf{s} \in \mathcal{M}^{N_\mathrm{B}}$ to complex \ac{IQ}-symbols. This corresponds to a simple lookup in a trainable embedding matrix $\mathbf{W} \in \mathbb{R}^{M \times 2}$ where $s_m$ ($m = 1,\,\ldots,\,{N_\mathrm{B}}$) determines the row of the matrix, of which the first column is the real and the second column the imaginary part of the resulting \ac{IQ}-sample. Here, $s_m$ is the $m$-th element of $\mathbf{s}$. After normalization, these \ac{IQ}-constellation symbols $\mathbf{c}$ shall be analyzed later to compare them with a classic 256-\ac{QAM} constellation. It follows an upsampling to simulation sampling rate $f_\mathrm{sim}$. The idea behind this is that the encoding \ac{DSP} shall be able to really learn a pulse-shaping, that we can visualize easily. With this, we try to make the encoder easier to understand, knowing that this increases the complexity but not its degree of freedom. Again, equivalently to a classic scheme, it follows a trainable pulse-shaper $\mathbf{F}$ that allows jointly learning of
\begin{itemize}
	\item a bandwidth-limitation to avoid information- or \ac{SNR}-loss by the \ac{LPF},
	\item a pulse-shape that propagates with little distortions, and
	\item a simple and low-level diversity coding, such that information may be distributed over time.
\end{itemize}
At last, the filtered signal undergoes an additional \ac{LPF} and a power-normalization with power $P$ to finally obtain the \ac{TX}-signal $\mathbf{x}$. The \ac{LPF} seems to be redundant as part of the pulse-shaping (see \ac{DAC} of the channel) and hence does not affect the overall performance of the system. Nevertheless, only the block-sequence \ac{LPF} first, and normalization second prevents the \ac{AE} from intentionally wasting power in the outside-bandwidth-regime\footnote{In case the normalization comes first, the \ac{LPF} would extract only a portion of the configured input power $P$.}, thus reducing the effective input power $P$, the influence of \ac{KNL} and hence manipulating the \ac{SNR}-definition. Further, this sequence of execution is necessary to preserve the gradient flow for frequencies outside $B_\mathrm{w}$, and hence to prevent the \ac{AE} from retaining high frequencies from the filter-weight-initialization. As the \ac{DAC}'s \ac{LPF} is fixed and as we do not want to add a normalization within the channel, we were forced to add an \ac{LPF} in the \ac{TX}-\ac{DSP}.

As shown in Fig. \ref{fig:ae-rx}, on the receiver side a \ac{DNN} with layers described in Tab. \ref{tab:rx-dnn} shall demap the received samples $\mathbf{y}_\mathrm{sam} \in \mathbb{C}^{N_\mathrm{B}}$ to the estimated sent symbol indices $\mathbf{\hat{s}} \in \mathcal{M}^{N_\mathrm{B}}$, where $\mathbf{y}_\mathrm{sam}$ again is the downsampled received signal $\mathbf{y}$. Therefore, it has not only access to the corresponding single sample, but also to its temporal neighbors. As shown in Fig. \ref{fig:ae-rx}, after downsampling the received signal, a sliding window of length $2N_\mathrm{adj}+1$ parallelizes the sequentially received samples to a matrix $\mathbf{Y}_\mathrm{adj}$. Hence, the input of the \ac{DNN} is a single row of $\mathbf{Y}_\mathrm{adj}$. With parameter $N_\mathrm{adj}$ we control the number of single-side neighbors.

\begin{table}[h]
	\caption{\ac{RX}-\ac{DNN}-configuration}
	\begin{center}
		\begin{tabular}{c|c|c}
			\textbf{Layer}&\textbf{Neurons}&\textbf{Activation function} \\
			\hline 
			&& \\
			1&2048&$\mathrm{elu}$ \\
			2-6&512&$\mathrm{elu}$ \\
			7&256&$\mathrm{softmax}$ \\
		\end{tabular}
		\label{tab:rx-dnn}
	\end{center}
\end{table}

Finally, the \ac{AE} is trained, such that the cross-entropy between the conditional \acp{PMF} of the labels and estimates is minimized. One can show that this optimization maximizes the spoken \ac{MI}.

\section{Results}\label{sec:results}

In the following we want to compare the performance between the conventional system and the \ac{AE}. We therefore start with a sanity check using two simplified channel models, one consisting only of \ac{AWGN}, denoted as $\mathcal{C}_\mathrm{A}$, and another with \ac{AWGN} and \ac{CD}, denoted as $\mathcal{C}_\mathrm{AD}$. In the beginning we set $N_\mathrm{adj}=0$ to ensure that all compensation has to be performed by the transmitter. As we know perfect transmission systems (achieving capacity limits) for those two channel-models, we can check whether the AE is able to also approximate those perfect solutions. The resulting SE is shown in Fig. \ref{fig:ae-performance}. As the \ac{AE} achieves the performance of the conventional system for $\mathcal{C}_\mathrm{A}$ and $\mathcal{C}_\mathrm{AD}$ this means that, \emph{jointly},
\begin{enumerate}
	\item a constellation that maximizes average distance between \ac{IQ}-symbols, and,
	\item a Nyquist-pulse-shaping, as well as
	\item a perfect \ac{CD}-compensation
\end{enumerate}
have been learned! Further, there is, of course, no gain compared to the conventional implementation. Nevertheless, this leads to the hypothesis that the AE is (at least in some cases) able to learn perfect solutions.

Applying the \ac{AE} to the channel-model $\mathcal{C}_\mathrm{ADN}$ (consisting of noise, \ac{CD} and \ac{KNL} as described in Section \ref{sec:system}) and starting with $N_\mathrm{adj}=0$ results in a small performance gain in terms of $SE$ compared to the conventional system only (see Fig. \ref{fig:ae-performance}). As we have chosen a classical communications structure we can now analyze the single blocks of the AE and hence try to find the source of this gain.

We first analyze the mapping of the \ac{AE} of Fig. \ref{fig:constellation}, which is indeed a more efficient constellation compared to a conventional 256-\ac{QAM} in case the $SE$ is calculated on uncoded symbols. Its circular shape can maximize the average distance between neighboring symbols. Furthermore, the density of symbols is lower in the high power than in the low power regime, tolerating higher \ac{KNL}-induced phase-shifts. This also matches with the results of \cite{jones2019end}. As both facts may already lead to a slightly higher performance we, nevertheless, analyze the surprising pulse shaping filter $\mathbf{F}$ of Fig. \ref{fig:ae-pulse-shapes-zero}, where a dark blue depicts the squared amplitude (power) of its impulse response, red depicts the received output pulse (after the channel) and dotted black a reference Nyquist-pulse. The same holds for its corresponding phase in blue, orange and gray.

In Fig. \ref{subfig:zero-amplitude-low} the signals' squared amplitudes for low powers ($P=-10\,\mathrm{dBm}$) are shown. Here, one can see that the \emph{received} output signal $\mathbf{y}$ matches with the well-known Nyquist-pulse perfectly. Hence, the AE shaped the input signal $\mathbf{x}$ such that it obtains orthogonal samples in time at the output, to not lose information by the implemented sampling (as $N_\mathrm{adj}=0$). Fig. \ref{subfig:zero-phase-low} further depicts the phase response of $\mathbf{F}$ that matches with the inverse of the well-known phase of \ac{CD} (wrapped parabolic) and, hence, a classic \ac{CD}-compensation, almost perfectly. The same holds for $P=0\,\mathrm{dBm}$ in Fig. \ref{subfig:zero-amplitude-mid} and \ref{subfig:zero-phase-mid}. At higher powers, where the AE-performance in terms of \ac{SE} follows the same curve as for the conventional system and especially at $P=10\,\mathrm{dBm}$, we experience a missing convergence of the \ac{AE} resulting in Fig. \ref{subfig:zero-amplitude-high} and \ref{subfig:zero-phase-high}. There, no sensible pulse-shapes were learned, that one can interpret. This may have two reasons, while we are not able to exclude the one or the other: First, the gradient may get lost such that optimization fails or, second, there is no proper constellation-pulse-shape-combination that works at high powers. Only by increasing $N_\mathrm{adj}$ we can see in Fig. \ref{fig:ae-performance} a significant performance gain from $1.32\,\mathrm{\sfrac{bits}{s \cdot Hz}}$ at $P=-10\,\mathrm{dBm}$ to $4.46\,\mathrm{\sfrac{bits}{s \cdot Hz}}$ at $P=0\,\mathrm{dBm}$, which increases with $N_\mathrm{adj}$ until it saturates at ca. $N_\mathrm{adj,opt}=20$. For higher values of $N_\mathrm{adj}$ there is no further gain (see Fig. \ref{fig:ae-performance} and $N_\mathrm{adj}=40$). Again, we conduct an analysis of the single blocks of the AE to search for the source of this gain. It turned out, that the constellations, generated by the mapping $\mathbf{W}$, again converge to the same constellation, shown in Fig. \ref{fig:constellation}. For low powers, there is no perfect convergence\footnote{As there is no gain for the AE being more precise or smoother than the channel's induced noise.} and hence, the constellation-diagram is not that symmetric, while still showing the same circular structure. We also interpret the learned pulse-shapes for $N_\mathrm{adj,opt}=20$ starting with their \ac{PSD} as shown in Fig. \ref{fig:psd-ae}.

Here, one can see for the input power with the highest gain (compared to classic systems) at $P=0\,\mathrm{dBm}$, that the \ac{AE} achieved a pulse-shaping that has much less spectral broadening by achieving a high \ac{DC}-offset (or a remaining \ac{CD}, respectively) at the output signal $\mathbf{y}_\mathrm{o}$. Unfortunately, it was not able to achieve the same for an even higher power $P=10\,\mathrm{dBm}$. Nevertheless, this leads to the hypothesis that the \ac{AE} tries to level the derivative of the signal's squared amplitude (power) and apply a phase modulation, to avoid \ac{KNL} in general. The corresponding filter shapes are shown in Fig. \ref{fig:ae-pulse-shapes-twenty}. All amplitudes at the three different power levels in Figs. \ref{subfig:twenty-amplitude-low}, \ref{subfig:twenty-amplitude-mid} and \ref{subfig:twenty-amplitude-high} do not provide any deeper insights. In contrast, the phase responses do: For low powers, one can still identify some wrapped parabolic shape of the learned pulses (see Fig. \ref{subfig:twenty-phase-low} and \ref{subfig:twenty-phase-mid}). Surprisingly, this is not only the case for the learned but also for the received phase. This means that the \ac{AE} left some \ac{CD} uncompensated before the pulse is transmitted over the channel. Unfortunately, this does not hold for high powers in Fig. \ref{subfig:twenty-phase-high}.

\section{Conclusion}\label{sec:conclusion}

Surprisingly, the learned pulse-shaping filter of the AE trained on a CD-only channel $\mathcal{C}_\mathrm{AD}$ matches the analytical solution for a CD-compensating filter almost perfectly, while the constellation is plausible. Hence, the AE is indeed able to learn a proper compensation of physical impairments while it can still be interpreted by choosing a classic communication structure, consisting of well-known but trainable blocks like a mapper or pulse-shaper. This gives rise to the hypothesis that an AE is also able of learning a compensation for a fiber model $\mathcal{C}_\mathrm{ADN}$ including also \ac{KNL} for which no closed form solution is known today.
In this work, we were able to achieve a gain of up to $4.46\,\mathrm{\sfrac{bits}{s\cdot Hz}}$ in terms of spectral efficiency by the AE's design for \ac{AWGN}, \ac{CD}, and \ac{KNL}. Nevertheless, it was not possible to compensate for the impairments over all input powers. To our believe, this is due to our purely linear design template of the AE's TX-DSP. Ongoing work focuses on the search for a nonlinear structure that is flexible enough to compensate better for nonlinear effects, while still being interpretable in terms of classic communcation signal processing.

\section*{Acknowledgement}

We want to thank Laurent Schmalen for all the exciting discussions and the chance for benefiting from his long experience in the field of machine learning for optical communications, especially in the beginning of this project.

\bibliographystyle{IEEEtran}
\bibliography{IEEEabrv,references}

\end{document}